\documentclass[prl,twocolumn,tightlines,showpacs,preprintnumbers,aps,floatfix,showpacs,amsmath,amssymb]{revtex4}
\usepackage{epsfig,amsmath,amssymb,bm,epsf,graphics}
\usepackage{dcolumn}

\def\lsim{\lower.5ex\hbox{$\; \buildrel < \over \sim \;$}}
\def\gsim{\lower.5ex\hbox{$\; \buildrel > \over \sim \;$}}

\newcommand{\be}{\begin{equation}}
\newcommand{\ee}{\end{equation}}
\newcommand{\ben}{\begin{eqnarray}}
\newcommand{\een}{\end{eqnarray}}

\begin{document}

\title{Analogous Hawking Radiation from Astrophysical Black Hole Accretion}
\author{Tapas K. Das}
\email{tapas@camk.edu.pl , tapas@astro.ucla.edu}
\affiliation{Nicolaus Copernicus Astronomical Center, Bartycka 18 00-716 Warsaw, Poland}
\affiliation{Division of Astronomy, University of California Los Angeles LA CA 90095-1562}

\date{\today}

\begin{abstract}
For the first time in literature, we provide an exact analytical
scheme for calculating the analogous Hawking temperature and surface
gravity for general relativistic accretion onto astrophysical black
holes. Our calculation may bridge the gap between the theory of
transonic astrophysical accretion and the theory of analogous Hawking
radiation, and may allow one to compare the properties of the two types
of horizons, electromagnetic and acoustic, in connection to the black
hole physics. We show that the domination of the analogous Hawking radiation
over the actual Hawking radiation may be a real astrophysical phenomena.
We also discuss the possibilities of the emergence of analogous white holes
around astrophysical black holes. Our calculation is general enough to
accommodate accreting black holes with any mass.
\end{abstract}


\maketitle
By providing an analogy between the propagation of light in 
a curved geometry and the acoustic propagation in a 
steady, inviscid, transonic barotropic fluid, it has been 
demonstrated \cite{unruh} that the representative fluid posses the
signature of an acoustic black hole (BH hereafter) embedded by an acoustic horizon
(AH hereafter) at its transonic point (where the fluid Mach number 
becomes unity).
This AH emits thermal radiations known as analogous Hawking radiation (AHR 
hereafter).
The temperature of such radiation is
called Analogous Hawking temperature, or acoustic Hawking temperature
(Unruh \cite{unruh} was the first to propose and compute such
temperature). 
Throughout our work, we use $T_{AH}$
to denote the analogous Hawking temperature, and $T_H$ to denote the
the actual Hawking temperature
where $T_H={\hbar{c^3}}/{8{\pi}{k_b}GM_{BH}}$,
$M_{BH}$ being the mass of the black hole. 
We use the words `analogous', `acoustic' and `sonic' synonymously
in describing the horizons or black holes.
Theory of AHR opens up the possibility to experimentally verify some 
basic features of BH physics by creating the sonic horizons in the 
laboratory. A number of works have been carried out to formulate the
condensed matter or optical analogous of event horizons
\cite{arti}. It is also 
expected that AHR may find important use in the field of Inflationary 
models, quantum gravity and sub-Planckian models of string theory
\cite{peran}. \\
\noindent
Since the publication of the seminal paper by Bondi in 1952
\cite{bondi}, the transonic 
behaviour of the accreting fluid onto compact astrophysical objects have 
been extensively studied in the astrophysics community (including the 
fast ever paper on curved acoustic geometry by Moncrief \cite{moncr}), and the 
pioneering work by Unruh in 1981 \cite{unruh}
initiated a substantial number of work
in the theory of AHR with diverse fields of application stated above. However, it is
quite astonishing that, any attempt is yet to be made to make a bridge
between these two categories of research, the astrophysical BH accretion
and the theory of AHR, by providing a self-consistent study of AHR for real 
astrophysical fluid flows. Since both the theory of transonic
astrophysical accretion (TTAA hereafter) and the theory of 
AHR stems out from almost 
exactly the same physics, the propagation of transonic fluid with 
acoustic disturbances embedded into it, we feel the pressing need to study 
AHR for relativistic transonic accretion onto astrophysical black 
holes and to compute $T_{AH}$ for such accretion. \\
We consider general 
relativistic (GR) spherically symmetric accretion of inviscid 
polytropic fluid onto a Schwarzschild BH of mass $M_{BH}$. We first formulate 
and solve the conservation equations for such accretion and then 
demonstrate that such flow may become transonic depending on initial boundary 
conditions. We then {\it analytically} calculate the {\it exact} location of
the acoustic horizon $r_h$ \cite{tapas} and calculate the 
quantities on AH which are required to compute the surface gravity and $T_{AH}$
for our model.
We then calculate $T_{AH}$ in {\it curved space time} for
{\it all possible}
real physical accretion solutions. \\
\noindent
The importance of our work, as we believe,
is not only lies in the fact that we, for the first time, {\it completely
analytically} compute the {\it exact} value of $T_{AH}$ for GR astrophysical fluid flow
in curved space time 
\footnote{Note that Unruh's original calculation of $T_{AH}$ \cite{unruh} assumes the 
positional constancy of sound speed, which may not always 
be a real physical situation.
This was modified to include the position dependent sound speed by 
Visser \cite{unruh} for flat space and by Bili$\acute{c}$ \cite{unruh} 
for curved space. However, none of the 
literatures provides the exact calculation of $r_h$ and AH related quantities
from the first principle, and hence the numerical value of $T_{AH}$ in {\it any}
existing literature (even for flat space, let alone the curve space) is 
obtained 
using 
approximately order of magnitude calculation only.}. 
Our calculation rather bridges the long standing gap between two 
apparently different school of thoughts sharing intrinsically the same 
physical origin for their work. And perhaps the most important point to 
note here that the accreting astrophysical BHs are the {\it only} 
real physical candidates for which both the horizons, actual (electromagnetic)
and analogous (sonic), may exist together {\it simultaneously}. Hence 
our application of AHR to TTAA may provide a novel procedure to compare 
the properties of these two kind of horizons in its most consistent way
i.e., by studying and comparing the behaviour of the {\it same} flow
close to these horizons. As
an illustrative example of such procedure, we also calculate 
and compare $T_{AH}$ and $T_H$ for the same BH and for same set of initial
boundary conditions, and show that for astrophysically interesting regions of
parameter space, $T_{AH}$ may well exceed $T_H$ which indicates that sometimes
AHR may be dominant over the actual Hawking radiation phenomena. Also we
find the possibility that an actual BH event horizon may be embedded by an
acoustic white hole. \\
We take the Schwarzschild radius $r_g=2GM_{BH}/c^2$, and scale the radial 
coordinate $r$ in units of $r_g$, any velocity in units of $c$, and all other
physical quantities are scaled accordingly. $G=c=M_{BH}=1$ is set. Mass of the 
accreting fluid is assumed to be quite less compared to $M_{BH}$ (which is 
reality for astrophysical BH accretion), so that the gravity field is 
controlled essentially by $M_{BH}$ only. Accretion is 
$\theta$ and $\phi$ symmetric and posses only radial inflow velocity. We 
concentrate on stationary solutions. Accretion is governed by the radial part
of the GR time independent Euler and continuity equations in Schwarzschild
metric. The conserved specific flow energy ${\cal E}$ (the relativistic 
analogue of Bernoulli's constant) along each stream line reads ${\cal E}=hu_t$ 
\cite{anderson} where
$h$ and $u_\mu$ are the specific enthalpy and the four velocity, which can be 
re-casted in terms of radial three velocity $u$ and polytropic sound speed $a$ to obtain:
$$
{\cal E}=\left[\frac{\gamma-1}{\gamma-\left(1+a^2\right)}\right]
\sqrt{\frac{1-\frac{1}{r}}{1-u^2}}
\eqno{(1)}
$$
The mass accretion rate ${\dot M}$ may be obtained by integrating the continuity
equation:
$$
{\dot M}=4{\pi}{\rho}ur^2\sqrt{\frac{r-1}{r\left(1-u^2\right)}}
\eqno{(2)}
$$
where $\rho$ is the proper mass density. $p=K\rho^\gamma$ equation of state is 
used (use of which is common in astrophysics
to describe relativistic accretion) to define another constant $\Xi$ as:
$$
\Xi=K^{\frac{1}{1-\gamma}}{\dot M}=4{\pi}{\rho}ur^2\sqrt{\frac{r-1}{r\left(1-u^2\right)}}
\left[\frac{a^2\left(\gamma-1\right)}{\gamma-\left(1+a^2\right)}\right]
\eqno{(3)}
$$
Simultaneous solution of eq. (1-3) provides the dynamical three velocity gradient 
at any radial distance $r$:
$$
\frac{du}{dr}=\frac{u\left(1-u^2\right)\left[a^2\left(4r-3\right)-1\right]}
{2r\left(r-1\right)\left(u^2-a^2\right)}=\frac{{\cal N}}{{\cal D}}
\eqno{(4)}
$$
The denominator 
of eq. (4) becomes zero at $r=1$ (actual horizon) and when $u=a$ (acoustic
horizon). To maintain the physical smoothness of the flow for any $r>1$, the numerator
${\cal N}$ and the denominator ${\cal D}$
of eq. (4) has to simultaneously vanish. Hence by making ${\cal N}$ = ${\cal D}$ = 0,
we get the so called sonic point conditions (value of dynamical and sound velocities
in terms of horizon location $r_h$) as:
$$
u_h=a_h=\sqrt{\frac{1}{4r_h-3}}
\eqno{(5)}
$$
Substitution of $u_h$ and $a_h$ in eq. (1) for $r=r_h$ provides:
$$
r_h^3+r_h^2\Gamma_1+r_h\Gamma_2+\Gamma_3=0
\eqno{(6)}
$$
where
$$
\Gamma_1=\left[\frac{2{\cal E}^2\left(2-3\gamma\right)+9\left(\gamma-1\right)}
         {4\left(\gamma-1\right)\left({\cal E}^2-1\right)}\right],~
$$
$$
\Gamma_2=\left[\frac{{\cal E}^2\left(3\gamma-2\right)^2-
          27\left(\gamma-1\right)^2}
          {32\left({\cal E}^2-1\right)\left(\gamma-1\right)^2}\right],~
\Gamma_3=\frac{27}{64\left({\cal E}^2-1\right)}
\eqno{(7)}
$$
Solution of eq. (6) provides the location of AH in terms of only two accretion parameters
$\{{\cal E},\gamma\}$, which is our two parameter input set to study the flow,
and hereafter
will be denoted by ${\cal P}_2$. We solve eq. (6) completely analytically 
by employing the Cardano-Tartaglia-del Ferro technique. We define:
$$
\Sigma_1=\frac{3\Gamma_2-\Gamma_1^2}{9},~
\Sigma_2=\frac{9\Gamma_1\Gamma_2-27\Gamma_3-2\Gamma_1^3}{54},~
\Psi=\Sigma_1^3+\Sigma_2^2,~
$$
$$
 \Theta={\rm cos}^{-1}\left(\frac{\Sigma_2}{\sqrt{-\Sigma_1^3}}\right)
\Omega_1=\sqrt[3]{\Sigma_2+\sqrt{\Sigma_2^2+\Sigma_1^3}},~
$$
$$
\Omega_2=\sqrt[3]{\Sigma_2-\sqrt{\Sigma_2^2+\Sigma_1^3}},~
\Omega_{\pm}=\left(\Omega_1\pm\Omega_2\right)
\eqno{(8)}
$$
so that the three roots for $r_h$ comes out to be:
$$
^1\!r_h=-\frac{\Gamma_1}{3}+\Omega_+,~
^2\!r_h=-\frac{\Gamma_1}{3}-\frac{1}{2}\left(\Omega_+-i\sqrt{3}\Omega_-\right),~
$$
$$
^3\!r_h=-\frac{\Gamma_1}{3}-\frac{1}{2}\left(\Omega_--i\sqrt{3}\Omega_-\right),
~\eqno{(9)}
$$
However, note that not all $^i\!r_h\{i=1,2,3\}$ would be real for all ${\cal P}_2$. It is
easy to show that if $\Psi>0$, only one root is real; if $\Psi=0$, all roots are
real and at least two of them are identical; and if $\Psi<0$, all roots are real 
and distinct. Selection of the real physical ($r_h$ has to be greater than unity) roots
requires
the following discussions. \\ 
Although we analytically calculate $r_h$ and other variables at $r_h$, there is no 
way that one can analytically calculate flow variable at any arbitrary $r$. One needs
to integrate eq. (6) only numerically to obtain the transonic profile of the flow for 
all range of $r$, starting from infinity and ending on to the actual event horizon. To do
so, one requires to set the appropriate limits on ${\cal P}_2$ to model the realistic situations
encountered in astrophysics. As ${\cal E}$ is scaled in terms of
the rest mass energy and includes the rest mass energy (${\cal E}=1$ corresponds to
rest mass energy), ${\cal E}<1$ corresponds to the negative energy accretion state where
radiative extraction of rest mass energy from the fluid is required. For such extraction 
to be made possible, accreting fluid has to
posses viscosity and other dissipative mechanism, which may violate the Lorenzian invariance.
On the other hand, although almost any ${\cal E}>1$ is mathematically allowed, large 
values of ${\cal E}$ represents flows starting from infinity 
with extremely high thermal energy, and ${\cal E}>2$ accretion represents enormously 
hot (thermal energy greater than the rest mass energy) initial ($r\rightarrow{\infty}$)
flow configurations, which is not properly conceivable in realistic astrophysical situations.
Hence we set $1{\lsim}{\cal E}{\lsim}2$. Now, $\gamma=1$ corresponds to isothermal accretion
where accreting fluid remains optically thin. This is the physical lower limit for 
$\gamma$. $\gamma<1$ is not realistic in accretion
astrophysics. On the other hand,
$\gamma>2$ is possible only for superdense matter 
with substantially large magnetic
field (which requires the accreting material to be governed by GR magneto-hydrodynamic 
equations, dealing with which
is beyond the scope of this paper) and direction dependent anisotropic pressure. We thus 
set $1{\lsim}\gamma{\lsim}2$ as well, so ${\cal P}_2$ has the boundaries
$1{\lsim}\{{\cal E},\gamma\}{\lsim}2$. However, one should note that the most preferred 
values of $\gamma$ for realistic BH accretion ranges from $4/3$ 
to $5/3$ \cite{frank}.\\
\noindent
Coming back to our solution for $r_h$, we find that for our preferred range of ${\cal P}_2$,
one {\it always} obtains $\Psi<0$ {\it only}. Hence the roots are always real and three real
unequal roots can be computed as:
$$
^1\!{{r}}_h=2\sqrt{-\Sigma_1}{\rm cos}\left(\frac{\Theta}{3}\right)
                  -\frac{\Gamma_1}{3},
^2\!{{r}}_h=2\sqrt{-\Sigma_1}{\rm cos}\left(\frac{\Theta+2\pi}{3}\right)
                  -\frac{\Gamma_1}{3},~
$$
$$
^3\!{{r}}_h=2\sqrt{-\Sigma_1}{\rm cos}\left(\frac{\Theta+4\pi}{3}\right)
                  -\frac{\Gamma_1}{3}
\eqno{(10)}
$$
We find that for {\it all} $1{\lsim}{\cal P}_2{\lsim}2$, $^2\!{{r}}_h$ becomes negative. We 
then find that $\{^1\!{{r}}_h,^3\!{{r}}_h\}>1$ for most values of our astrophysically 
tuned ${\cal P}_2$.
However, it is also found that $^3\!{{r}}_h$ does not allow steady physical flows to pass
through it, either $u$, or $a$, or both, becomes superluminal before the flow reaches
the actual event horizon, or the Mach number profile shows intrinsic fluctuations for 
$r<r_h$. We obtain this information by numerically integrating the 
complete flow profile passing through $^3\!{{r}}_h$. Hence it turns out that we need to
concentrate {\it only} on  $^1\!{{r}}_h$ for realistic astrophysical BH accretion. 
Both large ${\cal E}$ and large $\gamma$ enhances the thermal energy of the flow 
so that the 
accreting fluid acquires large radial velocity to overcome $a$ only at the 
close vicinity of the BH. Hence $r_h$ anti-correlates with ${\cal P}_2$. 
To obtain
$(dv/dr)$ and $(da/dr)$ on AEH, we apply L' Hospital's rule on 
eq. (4) and we have:
$$
\left(\frac{du}{dr}\right)_{r=r_h}=\Phi_{12}-\Phi_{123},~
$$
$$
\left(\frac{da}{dr}\right)_{r=r_h}=\Phi_4\left(\frac{1}{\sqrt{4r_h-3}}+\frac{\Phi_{12}}{2}
                            -\frac{\Phi_{123}}{2}\right)
\eqno{(11)}
$$
where
$$
\Phi_{12}=-\Phi_2/2\Phi_1,~
\Phi_{123}=\sqrt{\Phi_2^2-4\Phi_1\Phi_3}/2\Phi_1,~
$$
$$
\Phi_1=\frac{6r_h\left(r_h-1\right)}{\sqrt{4r_h-3}},~
\Phi_2=\frac{2}{4r_h-3}\left[4r_h\left(\gamma-1\right)-
        \left(3\gamma-2\right)\right]
$$
$$
\Phi_3=\frac{8\left(r_h-1\right)}{\left(4r_h-3\right)^{\frac{5}{2}}}
       [r_h^2\left(\gamma-1\right)^2-r_h\left(10\gamma^2-19\gamma+9\right)
$$
$$
       +\left(6\gamma^2-11\gamma+3\right)],~
\Phi_4=\frac{2\left(2r_h-1\right)-\gamma\left(4r_h-3\right)}
       {4\left(r_h-1\right)}
\eqno{(12)}
$$
GR description of the sonic geometry for 
generalized acoustic metric is provided by Bili$\acute{c}$ \cite{unruh}. Keeping in the mind that the 
surface gravity can be obtained by computing the norm of the gradient of the norm of the 
Killing fields evaluated at $r_h$, we use the methodology developed in 
Bili$\acute{c}$ \cite{unruh} to calculate 
the general relativistic expression for surface gravity, and $T_{AH}$ for our flow geometry
is obtained as:
$$
T_{AH}=\frac{\hbar{c^3}}{4{\pi}{k_b}GM_{BH}}
    \left[\frac{r_h^{\frac{1}{2}}\left(r_h-0.75\right)}
    {\left(r_h-1\right)^{\frac{3}{2}}}\right]
    \left[\frac{d}{dr}\left(a-u\right)\right]_{r=r_h}
\eqno{(13)}
$$
where the values of $r_h, (du/dr)_h$ and $(da/dr)_h$ is obtained using the system of
unit and scaling used in our paper. Stationarity approximation allows us not to discriminate
between acoustic apparent and event horizon. Note again, that, since $r_h$ and other quantities
appearing in eq (13) is {\it analytically} calculated as a function of
${\cal P}_2$, we provide an {\it exact analytical value} of the 
general relativistic analogous Hawking temperature for {\it all possible solutions} of an 
spherically accreting
astrophysical black hole system, which has never been done in the literature.
If $\sqrt{4r_h-3}(1/2-1/\Phi_4)(\Phi_{12}-\Phi_{123})>1$,
one {\it always} obtains $(da/dr<du/dr)_h$, which indicates the presence of the
acoustic white holes at $r_h$. This inequality holds good for
certain astrophysically relevant range of ${\cal P}_2$, see following discussions.
For a particular value of ${\cal P}_2$, we now define the quantity $\tau$ to be the ratio of
$T_{AH}$ and $T_H$, i.e., $\tau=T_{AH}/T_H$.
$\tau$ comes out to be independent of the mass of the black hole, 
which enables us to compare the properties of two kind of horizons (actual and acoustic) for
accreting BH with any mass. Our calculation is thus applicable for the widest 
possible range of BH candidates, starting from the primordial holes
to the super massive black holes
at galactic cecntres.\\
\begin{figure}
\vbox{
\vskip -2.5cm
\centerline{
\psfig{file=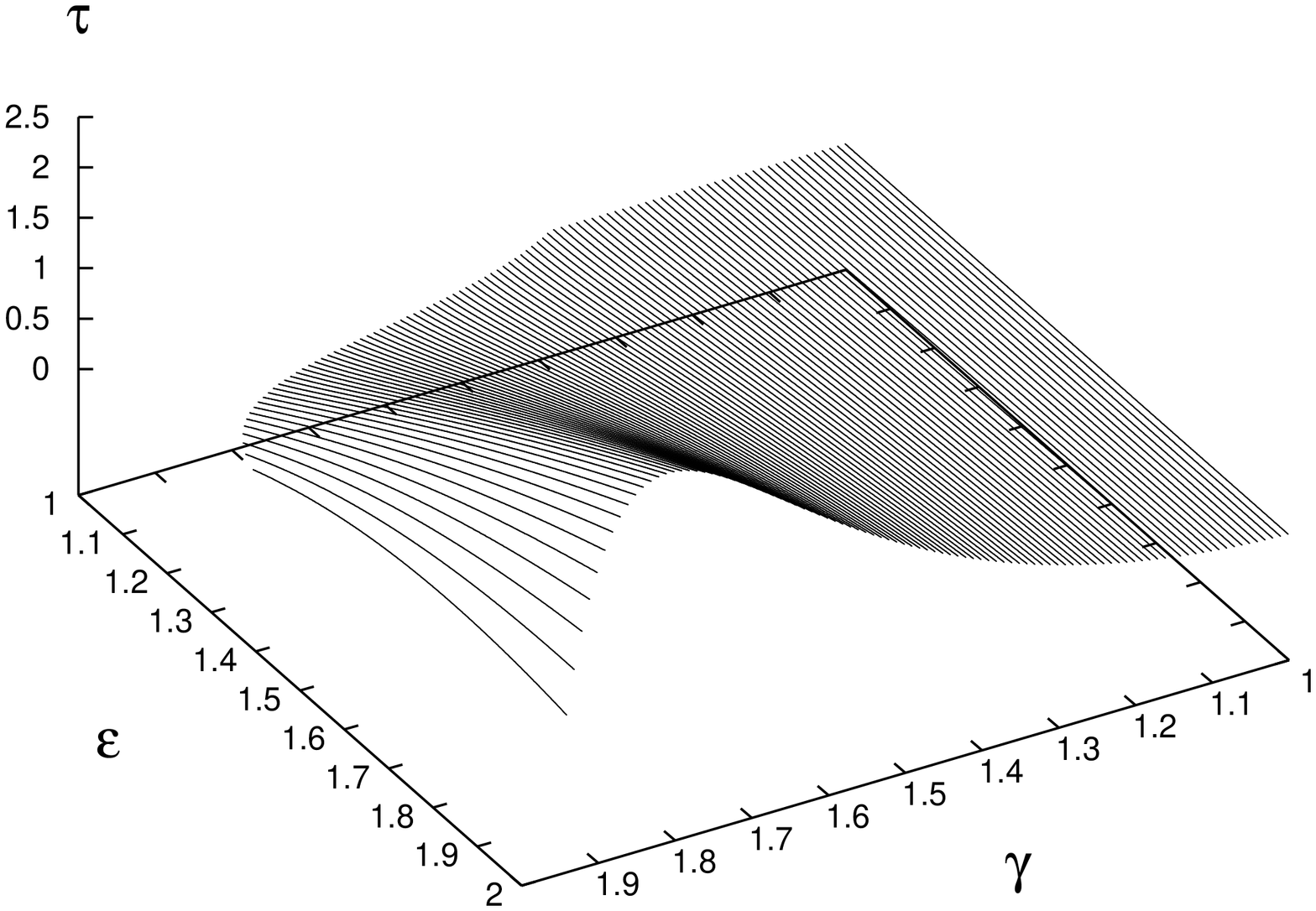,height=12cm,width=10cm,angle=270.0}}}
\noindent {{\bf Fig. 1:}
Variation of ratio of the analogous to the actual
Hawking temperature ($\tau$) on conserved specific energy (${\cal E}$)
and polytropic indices ($\gamma$) of the flow.}
\end{figure}
\begin{figure}
\vbox{
\vskip -4.5cm
\centerline{
\psfig{file=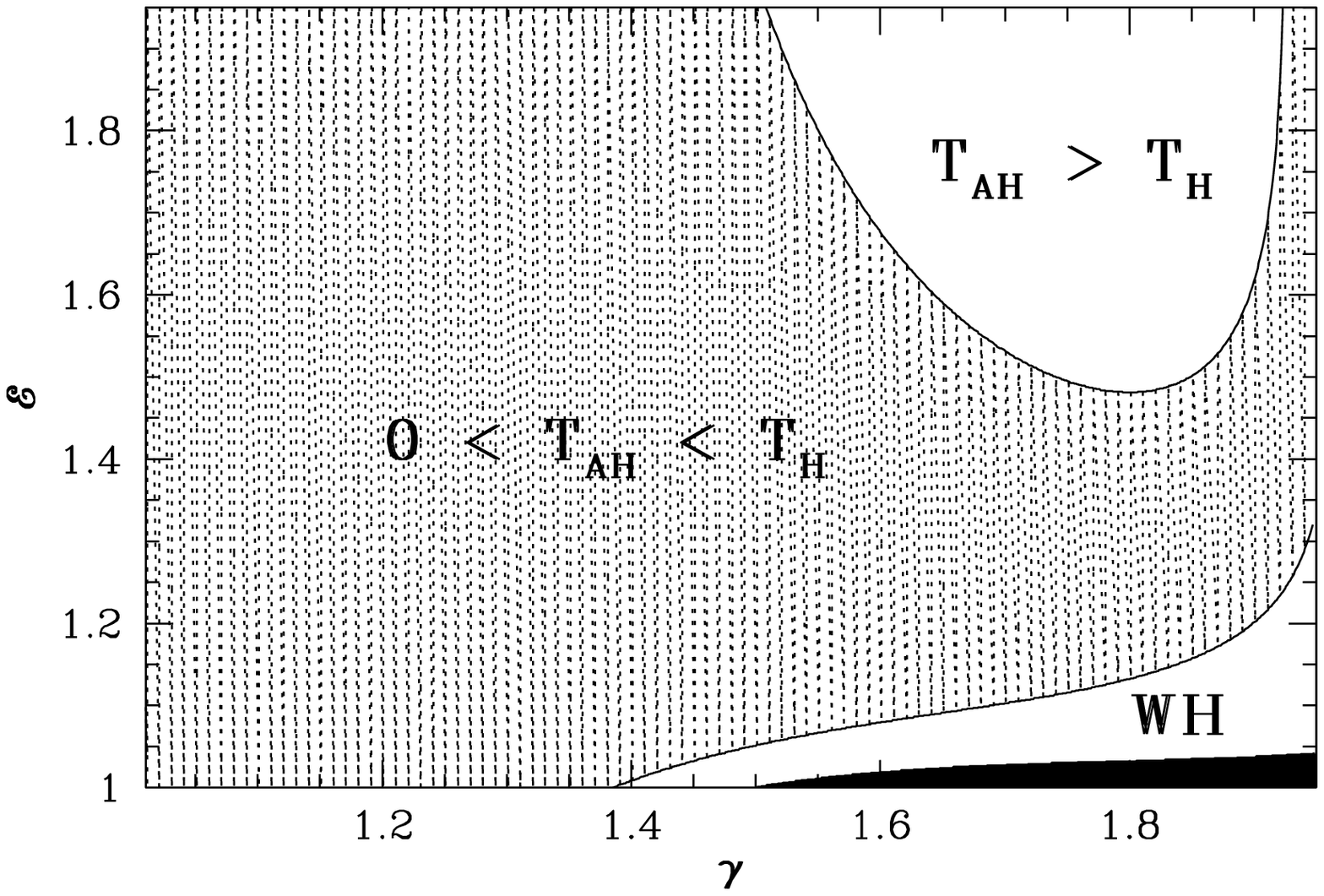,height=11cm,width=11cm,angle=0.0}}}
\noindent {{\bf Fig. 2:}
Parameter space classification for ${\cal P}_2$ for four different sets of values
of $T_{AH}$.}
\end{figure}
\noindent
Figure 1 shows the dependence of $\tau$ on ${\cal P}_2$. The dependence is highly nonlinear.
We plot $\tau$ coresponding to the black hole solutions only (solutions for which
$(da/dr)_h>(du/dr)_h$).
Note that $T_{AH}>T_H$ for
some ${\cal P}_2$. In figure 2., we scan the complete ${\cal P}_2$ space by depicting its
four important non-overlapping regions.
For accretion with ${\cal P}_2$ taken from the lightly shaded 
regions marked by $0<T_{AH}<T_H$, 
the Hawking radiation dominates over its analogous counterpart.
For high $\gamma$ high ${\cal E}$ flow (`hot' accretion),
AHR becomes the more dominant process compared to the actual Hawking radiation ($T_{AH}>T_H$).
For low ${\cal E}$ and intermediate/high $\gamma$, acoustic white holes appear
(white region marked by {\bf WH}, where ${da/dr<du/dr}$ at $r_h$). 
For ${\cal P}_2$ belonging to this region, one obtains outflow (outgoing solutions with Mach number 
increasing with increase of $r$) only. This has also been verified by obtaining the 
complete flow profile by integrating eq. (6).
The dark shaded 
region (lower right corner of the figure) represents ${\cal P}_2$ for which $r_h$ is comes out
to be real physical ($r_h>1$) but
$\Phi_{123}^2$ becomes negative, hence $(du/dr)_h$ and $(da/dr)_h$ are not real and $T_{AH}$
becomes imaginary. Note that both $T_{AH}>T_H$ and white hole regions are obtained even for 
$4/3<\gamma<5/3$, which are the values of $\gamma$ for most realistic flow of matter 
around astrophysical black holes.
{\it Hence we show that the domination of
the analogous Hawking radiation over the actual Hawking radiation and the emergence of 
analogous white holes are real astrophysical phenomena}. \\
However, the exact
astrophysical nature of 
the acoustic
white hole solutions are not quite clear to us at this moment. We did not perform formal stability
analysis on these solutions and it is possible that the white hole solutions are perhaps
unstable,
according to the recent proposal that the sonic white holes posses intrinsic instability \cite{leon}.
However, such detail investigations
are beyond the scope of this paper and we would like to defer it for our future work.
Note that calculations presented in this letter may be extended for astrophysical BH accretion
discs as well. Bili$\acute{c}$ \cite{unruh} studied the relativistic acoustic geometry for 2D axisymmetric space time
and one can use such formalism to study AHR for GR and post-Newtonian multi-transonic, rotating,
advective BH accretion discs in the similar way as is done in this letter. Such calculations
are in progress and will be presented elsewhere.
\begin{acknowledgements}
I am deeply indebted to Neven Bili$\acute{c}$ for his constant help during my process
of understanding the analogous Hawking radiation in curved geometry, 
and to William Unruh and Ted Jacobson 
for their insightful suggestions provided after going through
the first draft of this manuscript, which have been very useful to modify the 
manuscript to this present form. I also
acknowledge stimulating discussions with 
Paul Wiita, Matt Visser and Ralf Sch$\ddot{\rm u}$tzhold. 
\end{acknowledgements}


\begin{thebibliography}{99}
\bibitem{unruh}
W. G. Unruh, Phys. Rev. Lett. {\bf 46}, 1351 (1981);
T. A. Jacobson, Phys. Rev. D {\bf 44}, 1731 (1991);
W. G. Unruh, Phys. Rev. D {\bf 51}, 2827 (1995);
M. Visser, Class. Quant. Grav. {\bf 15}, 1767 (1998);
T. A. Jacobson, Prog. Theor. Phys. Suppl. {\bf 136}, 1 (1999);
N. Bili$\acute{c}$, Class. Quant. Grav. {\bf 16}, 3953 (1999).
\bibitem{arti}
See M. Novello, M. Visser, and G. E. Volovik (editors),
{\it Artificial Black Holes} (World Scientific, Singapore, 2002) for 
detail review.
\bibitem{peran}
R. Parentani, International Journal of Modern Physics A. {\bf 17},
2721 (2002) and references therein.
\bibitem{bondi}
H. Bondi, Monthly Notices of the Royal Astronomical Society. {\bf 112},
195 (1952).
\bibitem{moncr}
V. Moncrief, Astrophysical Journal. {\bf 235}, 1038 (1980).
\bibitem{tapas}
See T. K. Das, Astrophysical Journal. {\bf 577},
880 (2002); T. K. Das, J. K. Pendharkar, and S. Mitra, 
Astrophysical Journal. {\bf 592}, 1078 (2003);
T. K. Das, Monthly Notices of the Royal Astronomical Society.
{\bf 349}, 375 (2004) and references therein for the details of
semi-numerical sonic point analysis of more complex type of accretion
flows with multiple sonic surfaces.
\bibitem{anderson}
M. Anderson, Monthly Notices of the Royal Astronomical Society. {\bf 239},
19, (1989).
\bibitem{frank}
J. Frank, A. King, and D. Raine, {\it Accretion Power in Astrophysics}
3rd Edition (Cambridge University Press, Cambridge, UK, 2002).
\bibitem{leon}
U. Leonhardt, T. Kiss, and P. Ohberg, {\it Intrinsic instability of sonic white holes}.
gr-qc/0211069 (2002).






\end{thebibliography}
\end{document}